# Unbreakable $\mathcal{PT}$ symmetry of solitons supported by inhomogeneous defocusing nonlinearity

Yaroslav V. Kartashov,[1,2,*] Boris A. Malomed,[3] and Lluis Torner[1]

[1]*ICFO-Institut de Ciencies Fotoniques, and Universitat Politecnica de Catalunya, Mediterranean Technology Park, 08860 Castelldefels (Barcelona), Spain*
[2]*Institute of Spectroscopy, Russian Academy of Sciences, Troitsk, Moscow Region, 142190, Russia*
[3]*Department of Physical Electronics, School of Electrical Engineering, Faculty of Engineering, Tel Aviv University, Tel Aviv 69978, Israel*

We consider bright solitons supported by a symmetric inhomogeneous defocusing nonlinearity growing rapidly enough toward the periphery of the medium, combined with an antisymmetric gain-loss profile. Despite the absence of any symmetric modulation of the linear refractive index, which is usually required to establish a $\mathcal{PT}$ symmetry in the form of a purely real spectrum of modes, we show that the $\mathcal{PT}$ symmetry is *never* broken in the present system, and that the system *always* supports *stable* bright solitons, fundamental and multi-pole ones. Such phenomenon is connected to non-linearizability of the underlying evolution equation. The increase of the gain-losses strength results, in lieu of the $\mathcal{PT}$ symmetry breaking, in merger of pairs of different soliton branches, such as fundamental and dipole, or tripole and quadrupole ones. The fundamental and dipole solitons remain stable for all values of the gain-loss coefficient.

The high current interest in $\mathcal{PT}$-symmetric systems with complex potentials is partially motivated by the remarkable behavior of the corresponding linear spectrum, which remains purely real until the strength of the imaginary part of the potential attains a certain $\mathcal{PT}$-symmetry breaking threshold, above which it becomes complex [1]. The effect has been experimentally observed in optics, where the similarity of the paraxial evolution equation, governing the propagation of light beams in media with an even spatial profile of the refractive-index modulation and an odd gain-loss profile, with the Schrödinger equation governing the evolution of the quantum-mechanical wave function in a complex potential, enables visualization of the $\mathcal{PT}$ symmetry and its breakup at a critical point [2]. While eigenmodes of linear $\mathcal{PT}$-symmetric potentials are well understood [1,3,4], the evolution of nonlinear excitations in them remains a subject of active research. In particular, the properties of solitons and discrete nonlinear modes have been studied in free-standing $\mathcal{PT}$-symmetric waveguides [5], couplers [6-9], oligomers [10,11], periodic lattices with [12] and without [13-17] transverse refractive-index gradients in truncated lattices [18], among other settings. An especially interesting situation occurs when the underlying evolution equations contain only nonlinear $\mathcal{PT}$-symmetric terms [19,20], or mixed linear-nonlinear lattices [21-23].

As mentioned above, a generic property of the $\mathcal{PT}$-symmetric structures is that they support stable excitations only if the spectrum of the associated linear system is real, i.e., the symmetry is not broken. As a result, the stability domain of nonlinear excitations, if defined in terms of the gain-loss strength, often coincides with the domain of the unbroken $\mathcal{PT}$-symmetry in the respective linear system. However, systems may be prepared to be *non-linearizable*, i.e., the nonlinear terms in the underlying evolution equation cannot be omitted even for decaying tails of localized nonlinear excitations. Under such conditions, no direct link can be drawn between the spectra of the nonlinear system and its linear counterpart.

In this paper we address that case in a system with an odd gain-loss profile and defocusing nonlinearity whose local strength grows toward the periphery. In the absence of gain and loss, such system supports bright solitons in all three dimensions [24-28]. Existence of bright solitons in spite of the self-defocusing nature of the nonlinearity is at first counterintuitive, but actually it is a consequence of the nonlinearizability of the nonlinear Schrödinger equation on the solitons tails. Here we show that a system of this type, with an even profile of the growing nonlinearity, supports stable bright solitons even in the presence of the odd ($\mathcal{PT}$-symmetric) gain-loss profile *without* any spatial modulation of the linear refractive index, which is required for supporting the real spectrum in usual $\mathcal{PT}$-symmetric systems. While, without the rapidly growing defocusing nonlinearity, the $\mathcal{PT}$ symmetry of the system considered here is always *broken*, in the presence of the nonlinearity modulation the symmetry may be said to become *unbreakable*, as it holds at arbitrarily large strengths of the balanced gain and loss. Recently, unbreakable symmetry was demonstrated for a dimer, but it was a very special case of a $\mathcal{PT}$-symmetric Hamiltonian system [29].

We address the propagation of the laser beam along the $\xi$-axis of a medium with a transverse modulation of the gain-loss and defocusing nonlinearity, obeying the paraxial nonlinear Schrödinger equation for scaled amplitude $q$ of the electromagnetic field:

$$i\frac{\partial q}{\partial \xi} = -\frac{1}{2}\frac{\partial^2 q}{\partial \eta^2} + \sigma(\eta)q|q|^2 + iR(\eta)q, \qquad (1)$$

where the propagation distance $\xi$ is normalized to the diffraction length $kx_0^2$; the transverse coordinate $\eta$ is normalized to the characteristic transverse scale $x_0$; the function $\sigma(\eta) > 0$, which is assumed to be even, describes the profile of a defocusing nonlinearity whose strength grows as $\eta \to \pm\infty$; and the function $R(\eta)$, assumed to be odd, stands for the gain-loss profile. Following Refs. [24,25], we adopt the following nonlinearity and gain-loss profiles:

$$\sigma = (\sigma_0 + \sigma_2 \eta^2) \exp(\alpha \eta^2), \quad (2)$$
$$R = \beta \eta \exp(-\gamma \eta^2),$$

However, we note that the necessary and sufficient condition for the existence of bright solitons with power (or norm), $U = \int_{-\infty}^{\infty} |q|^2 \, d\eta$, sustained by a growing defocusing nonlinearity, is weaker. Namely, $\sigma(\eta)/|\eta| \to \infty$ at $|\eta| \to \infty$ [24]. Nevertheless, here we use the steep modulation profiles described by Eq. (2) because they create tightly localized solitons, which are convenient for the numerical and analytical considerations alike. The odd profile of $R(\eta)$ accounts for the mutually balanced attenuation of the field at $\eta < 0$, and amplification at $\eta > 0$. Nonlinearity and gain landscapes are depicted in Fig. 1(j) for $\sigma_0 = 1$, $\sigma_2 = 0$, and $\alpha = \gamma = 1/2$. The model gives rise to bright soliton solutions in the form $q(\eta, \xi) = [w_r(\eta) + i w_i(\eta)] \exp(ib\xi)$. Despite the dissipative character of the system, such solitons form continuous families, like in conservative media, parameterized by the propagation constant $b < 0$. The complex form of the solitons gives rise to the intrinsic currents, $j = w^2 d\phi/d\eta$, where $w \exp(i\phi) = w_r + i w_i$, with $w_r$ and $w_i$ being real and imaginary parts of solution, respectively.

We have found multiple soliton families, including fundamental [Figs. 1(a),(b)], dipole [Figs. 1(c),(d)], tripole [Figs. 1(e),(f)], and quadrupole [Figs. 1(g),(h)] soliton solutions, and even more complex states. Their multi-pole structure is readily visible mostly for small values of $\beta$, when soliton solutions are close to their conservative counterparts. Increasing $\beta$ leads to a notable transformations of the field-amplitude distributions, as shown in the top and bottom lines of Fig. 1. The intrinsic currents rapidly grow when $\beta$ increases, but the profile of $j(\eta)$ remains bell-shaped even for solitons with a large number of humps [Fig. 1(i)].

Next we vary the gain-loss strength, $\beta$, for a fixed propagation constant, in order to study the impact of the imaginary potential in Eq. (1) on the soliton properties. Without nonlinearity modulation, the $\mathcal{PT}$-symmetry in Eq. (1) is broken, as the real part of the linear potential is zero. However, in what follows we show that the nonlinear *pseudo-potential* [30], $\sigma(\eta)|q|^2$, does sustain the $\mathcal{PT}$-symmetry. A key insight comes from the observation that Eqs. (1) and (2) with $\gamma = 0$ admit the exact soliton solution $q = (2\sigma_2)^{-1/2} \alpha \exp(ib\xi - i\beta\eta/\alpha - \alpha\eta^2/2)$, with propagation constant $b = -[\alpha + \sigma_0 \alpha^2/\sigma_2 + (\beta/\alpha)^2]/2$, which can be found for *arbitrarily large* values of the gain-loss coefficient $\beta$. Moreover, this solution can be completely stable (the stability was checked at $\sigma_2 = \alpha = \beta = 1/2$), indicating on the conservation of the $\mathcal{PT}$-symmetry in the presence of pseudo-potential.

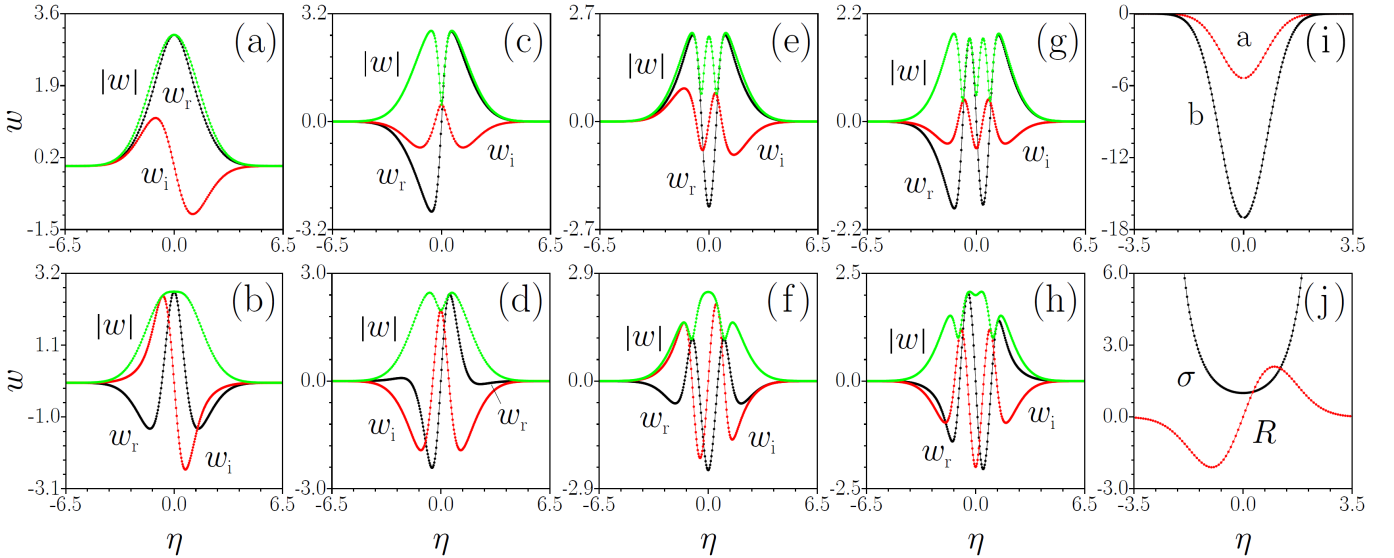

Fig. 1. (Color online) Profiles of fundamental (a),(b), dipole (c),(d), tripole (e),(f), and quadrupole (g),(h) solitons at $b = -10$. Panels (a),(c) correspond to $\beta = 0.55$, in (b),(d) $\beta = 1.98$, in (e),(g) $\beta = 1.04$, while in (f),(h) $\beta = 3.47$. Panel (i) shows currents in fundamental solitons from (a) and (b). Panel (j) shows nonlinearity and gain-loss landscapes for $\beta = 3.47$.

The soliton families are characterized by the dependences of the total power on $\beta$, which are depicted in Figs. 2(a) and 2(b) for the four types of modes shown in Fig. 1. Here subscripts f, d, t, and q refer to the families of fundamental, dipole, tripole, and quadrupole solitons, respectively. A first noteworthy result is that different soliton families, with completely different internal structure when $\beta = 0$ merge at different critical values of $\beta$, for a fixed propagation constant, $b$. Specifically, the family of fundamental solitons merges with the family of dipoles, the family of tripoles merges with the quadrupoles, and so on. We observed that the value, $\beta_{\mathrm{upp}}$, at which merging occurs increases with the order of the soliton families, as visible by comparing $\beta_{\mathrm{upp}}$ values for the dipole "d" and quadrupole "q" families in Fig. 2(c). Close to the merger point, the distributions of the absolute value of the field amplitude for solitons belonging to the merging branches are nearly indistinguishable, cf. Figs. 1(f) and 1(h). The numerical calculations also reveal that the soliton power decreases when the number of lobes in the soliton profiles increases for given values of $b$ and $\beta$, and that the largest gain-loss strength, $\beta_{\mathrm{upp}}$, at which different families merge, monotonously increases with $|b|$ for all types of solutions considered here [Fig. 2(c)].

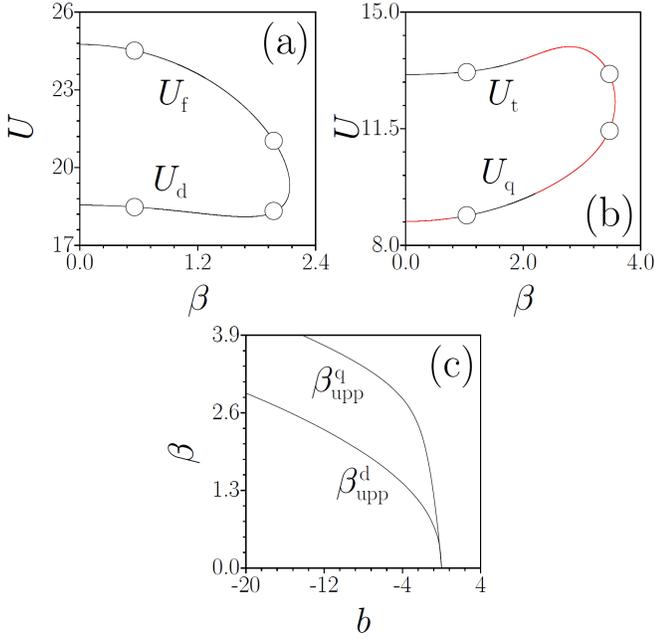

Fig. 2. (Color online) Power $U$ versus gain-losses strength $\beta$ for fundamental and dipole solitons (a), as well as for tripole and quadrupole solitons (b). Circles correspond to solutions shown in Fig. 1. The fundamental and dipole families merge at $\beta_{\text{upp}}^{\text{f,d}} = 2.135$, while the tripole and quadrupole ones merge at $\beta_{\text{upp}}^{\text{t,q}} = 3.565$. Black and red lines correspond to stable and unstable subfamilies, respectively. (c) Largest $\beta$ below which various soliton families exist, as a function of propagation constant $b$ (at $\beta \to 0$, the curves originate from $b = 0$, which correspond to solitons with zero amplitude).

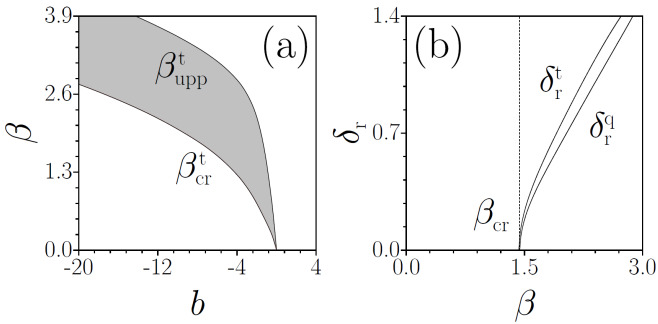

Fig. 3. (a) Domains of stability (white) and instability (shaded) for tripole solitons in the plane of $(b, \beta)$. (b) The real part of the instability growth rate vs. $\beta$ at $b = -5$ for tripoles and quadrupoles.

Therefore, despite the absence of any linear-refractive-index modulation profile in Eq. (1), the $\mathcal{PT}$-symmetry, supported by the nonlinear pseudo-potential, is indeed *not broken* in the present system, because the fundamental and higher-order stable solitons are found for arbitrarily large $\beta$, at sufficiently large values of $|b|$. Below we will show that such soliton solutions can also be *stable* at arbitrarily large $\beta$ values – a clear indication of unbreakable symmetry in the present setting.

On physical grounds, self-localization around the minimum of $\sigma(\eta)$ is provided by the growth of the local strength of the defocusing nonlinearity at $\eta \to \pm\infty$ at any rate faster than $|\eta|$, as mentioned above. It is thus not possible to drop the nonlinear term in Eq. (1) even for the vanishing tails of the solution, due to the rapid growth of $\sigma(\eta)$ [24,25]. Accordingly, the nonlinear pseudo-potential $\sigma(\eta)|q|^2$ in Eq. (1) provides the balance with the odd imaginary potential. This is in contrast with previously considered $\mathcal{PT}$-symmetric systems with uniform nonlinearity, where nonlinear terms may be dropped far from the soliton center and the conditions for maintaining the $\mathcal{PT}$-symmetry are determined by the linear terms in the evolution equation.

To test the dynamic stability of the soliton families we studied the evolution of the perturbed solutions in the form $q = [w_r + iw_i + u\exp(\delta\xi) + iv\exp(\delta\xi)]\exp(ib\xi)$, with $|u, v| \ll |w_r, w_i|$. Substitution of this field into Eq. (1) and linearization leads to the eigenvalue problem

$$\delta u = -\frac{1}{2}\frac{\partial^2 v}{\partial \eta^2} + Rv + \sigma[2uw_iw_r + v(3w_i^2 + w_r^2)] + bv,$$
$$\delta v = +\frac{1}{2}\frac{\partial^2 u}{\partial \eta^2} + Rv - \sigma[2vw_iw_r + u(3w_r^2 + w_i^2)] - bu,$$ (3)

that we solved numerically to elucidate the growth rates $\delta$ of the perturbations. We found that the families of fundamental and dipole soliton are fully stable for any value of propagation constant $b$ (even in small power limit $U \to 0$) and gain/loss strength $\beta$ within the existence domain of such states [Fig. 2(c)]. Instabilities emerge only for some of the tripole and quadrupole solutions. The unstable solutions are depicted in red in Fig. 2(b). In particular, tripoles are stable for $0 < \beta < \beta_{\text{cr}}$, and unstable for $\beta_{\text{cr}} < \beta < \beta_{\text{upp}}$. The corresponding critical gain-loss strength $\beta_{\text{cr}}$ is shown, as a function of propagation constant $b$, in Fig. 3(a), along with the $\beta_{\text{upp}}(b)$ curve. One observes that the stability region notably broadens when the propagation constant decreases. The shape of the stability area for quadrupole solitons is similar at small values of $|b|$, but additional instability windows appear at large values of $|b|$. In general, tripole and quadrupole solutions become unstable at different critical values of the gain-loss strengths, but a situation is possible when $\beta_{\text{cr}}$ nearly coincide for two families. Typical dependences of the instability growth rates as a function of $\beta$ for tripole and quadrupole solutions families are shown in Fig. 3(b). Examples of the stable evolution of the four soliton families considered here are displayed in Figs. 4(a)-4(c) and 4(e). Even in the presence of strong perturbations, such solitons propagate for unlimited distances without visible distortions. In contrast, unstable tripoles and quadrupoles feature progressively swinging amplitude oscillations in Figs. 4(d) and 4(f).

In summary, we have introduced the first example of a nonlinear $\mathcal{PT}$-symmetric system with an *unbreakable symmetry*. Namely, for sufficiently large propagation constants, soliton solutions of the system remain stable for arbitrarily large values of the gain-loss strength. In addition, families of fundamental and dipole solitons are completely stable. On physical grounds, in this system the spatially odd gain-loss distribution is balanced not by an even refractive-index profile, but by the pseudo-potential induced by the spatially growing strength of the defocusing nonlinearity.

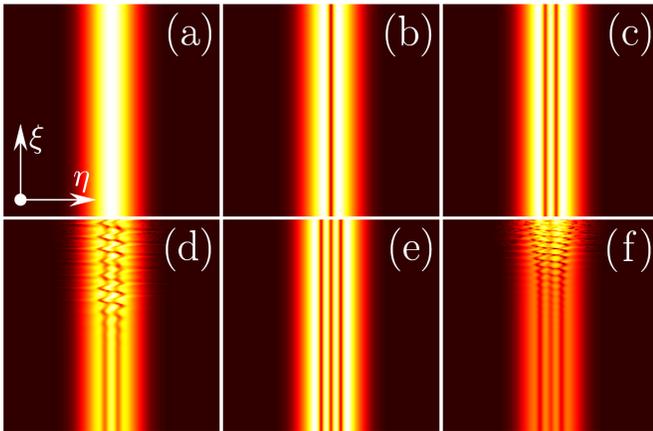

Fig. 4. (Color online) Stable propagation of fundamental (a) and dipole (b) solitons at $\beta = 0.55$, a tripole (c) and quadrupole (e) at $\beta = 1.04$. Instability development for (d) a tripole at $\beta = 2.10$ and (f) quadrupole at $\beta = 2.30$. In all the cases, $b = -10$.